\documentclass[preprint,aps,prd,showpacs,nofootinbib]{revtex4}
\usepackage{mathrsfs}
\usepackage{amsmath}
\usepackage{graphicx}
\usepackage{subfigure}

\newcommand{\bea}{\begin{eqnarray}}
\newcommand{\eea}{\end{eqnarray}}
\newcommand{\beq}{\begin{equation}}
\newcommand{\eeq}{\end{equation}}

\def\k{{\vec k}}

\def\/{\over}

\begin{document}

\title{Do static atoms outside a Schwarzschild \\ black hole spontaneously excite?}
\author{ Hongwei Yu and Wenting Zhou  }
\affiliation{Department of Physics and Institute of  Physics,\\
Hunan Normal University, Changsha, Hunan 410081, China }

\begin{abstract}

The spontaneous excitation of a two-level atom held static outside a
four dimensional Schwarzschild black hole and in interaction with a
massless scalar field in the Boulware, Unruh and Hartle-Hawking
vacuum is investigated and the contributions of the vacuum
fluctuations and radiation reaction to the rate of change of the
mean atomic energy are  calculated separately. We find that for the
Boulware vacuum, the spontaneous excitation does not occur and the
ground state atoms are stable, while the spontaneous emission rate
for excited atoms in the Boulware vacuum, which is well-behaved at
the event horizon, is not the same as that in the usual Minkowski
vacuum. However, both for the Unruh vacuum and the Hartle-Hawking
vacuum, our results show that the atom  would spontaneously excite,
as if there were  an outgoing thermal flux of radiation or as if it
were in a thermal bath of radiation at a proper temperature which
reduces to the Hawking temperature  in the spatial asymptotic
region, depending on whether the scalar field is in the Unruh or
Hartle-Hawking vacuum.

\end{abstract}
\maketitle

\baselineskip=16pt

\section{Introduction}

Spontaneous emission is one of the most important features of atoms
 and so far mechanisms such as vacuum
fluctuations \cite{Welton48, CPP83}, radiation reaction
\cite{Ackerhalt73}, or a combination of them \cite{Milonni88} have
been put forward to explain why spontaneous emission occurs. The
ambiguity in physical interpretation arises because of the freedom
in the choice of ordering of commuting operators of the atom and
field in a Heisenberg picture approach to the problem. The
controversy was resolved when Dalibard, Dupont-Roc and
Cohen-Tannoudji(DDC) \cite{Dalibard82,Dalibard84} proposed a
formalism which distinctively separates the contributions of vacuum
fluctuations and radiation reaction by demanding a symmetric
operator ordering of atom and field variables.  The DDC formalism
has recently been generalized to study the spontaneous excitation of
uniformly accelerated atoms in interaction with vacuum fluctuations
of  scalar and electromagnetic fields in a flat
spacetime~\cite{Audretsch94,H. Yu,ZYL06,YuZ06,ZYu07}, and these
studies show that when an atom is accelerated, the delicate balance
between vacuum fluctuations and radiation reaction that ensures the
ground state atom's stability in vacuum is altered, making possible
the transitions to excited states for ground-state atoms even in
vacuum.

Inspired by an equivalence principle-type argument, i.e., the same
accelerated atoms are seen by comoving observers as static ones in a
uniform ``gravitational field",  one may wonder what happens if an
atom is held static in a curved spacetime,  such as that of a black
 hole, for example. Do static atoms spontaneously excite outside a
 black hole? and if it does, will the excitation rate be
 what one expects assuming the existence of Hawking radiation from black holes?
 Answer to these questions may reveal relationship between  the Hawking radiation and the
 spontaneous excitation of atoms outside a black hole,
 and thus
 provide alternative derivation of Hawking radiation.
When we move to study the spontaneous excitation of static atoms
interacting with vacuum fluctuations of quantum fields in a curved
spacetime, a delicate issue then arises as to how the vacuum state
of the quantum fields is determined. Normally, a vacuum state is
associated with non-occupation of positive frequency modes. However,
the positive frequency of field modes are defined with respect to
the time coordinate. Therefore, to define positive frequency, one
has to first specify a definition of time. In a spherically
symmetric black hole background, one definition is the Schwarzschild
time, $t$ , and it is a natural definition of time in the exterior
region. The vacuum state, defined by requiring normal modes to be
positive frequency with respect to the Killing vector $\partial/
\partial t$ with respect to which the exterior region is static, is
called the Boulware vacuum. Other possibilities that have been
proposed are the Unruh vacuum~\cite{Unruh} and the Hartle-Hawking
vacuum~\cite{Hartle-Hawking}. The Unruh vacuum is defined by taking
modes that are incoming from $\mathscr{J}^-$ to be positive
frequency with respect to $\partial/ \partial t$, while those that
emanate from the past horizon are taken to be positive frequency
with respect to the Kruskal coordinate $\bar u$, the canonical
affine parameter on the past horizon.  The Hartle-Hawking vacuum, on
the other hand, is defined by taking the incoming modes to be
positive frequency with respect to $\bar v$, the canonical affine
parameter on the future horizon, and outgoing modes to be positive
frequency with respect to $\bar u$.  The calculations of the values
of physical observables, such as the expectation values of the
energy-momentum tensor and the response rate of an Unruh detector in
these vacuum states, have yielded the following  physical
understanding:

 (i) The Boulware vacuum corresponds to our familiar
concept of a vacuum state at large radii, but is problematic in the
sense that the expectation value of the energy-momentum tensor,
evaluated in a free falling frame, diverges at the horizon.

(ii) The Unruh vacuum is the vacuum state that best approximates the
state that would obtain following the gravitational collapse of a
massive body, since in the spatial asymptotic region, it corresponds
to an outgoing flux of black-body radiation at the Hawking
temperature.

(ii) The Hartle-Hawking state, however, does not correspond to our
usual notion of a vacuum, as it has thermal radiation incoming to
the black hole from infinity and describes a black hole in
equilibrium with a sea of thermal radiation.

In the current paper, we would like to apply the DDC formalism to
study the spontaneous excitation of a static two-level atom  outside
a 4-dimensional Schwarzschild black hole in interaction with
massless quantum scalar fields in all the above three vacuum states,
aiming to answer the question of whether a static atom outside a
black hole spontaneously excite. We also hope to gain more insights
into the physical meaning of the vacuum states proposed so far in
the black hole spacetime, as well as to reveal relationship between
the Hawking radiation and spontaneous excitation of atoms. Let us
note that recently, we have already studied the spontaneous
excitation of a static two-level atom interacting with massless
scalar fields in both the Unruh vacuum and the Hartle-Hawking vacuum
outside a 1+1 dimensional Schwarzschild black hole and found that
the atom spontaneously excites as if there is thermal radiation at
the Hawking temperature emanating from the black
hole~\cite{YuZhou07}.


\section{General formalism}

Let us consider a two-level atom in interaction with a quantum real
massless scalar field outside a Schwarzschild black hole. The metric
of the spacetime can be written in terms of the Schwarzschild
coordinates as

\begin{equation}
ds^2= -\bigg(1-{2M\over r}\bigg)\;dt^2+\bigg(1-{2M\over
r}\bigg)^{-1}\;dr^2+r^2\,(d\theta^2+\sin^2\theta\,d\varphi^2)\;,
\end{equation}
wher $M$ is the mass of the black hole. Without loss of generality,
we assume a pointlike two-level atom on a stationary space-time
trajectory $x(\tau)$, where $\tau$ denotes the proper time on the
trajectory. The stationarity of the trajectory guarantees the
existence of stationary atomic states, $|+ \rangle$ and $|-
\rangle$, with energies $\pm{1\/2}\omega_0$ and a level spacing
$\omega_0$.  The atom's Hamiltonian which controls the time
evolution with respect to $\tau$ is given, in Dicke's notation
\cite{Dicke},  by
 \beq H_A (\tau)
=\omega_0 R_3 (\tau)\;,  \label{atom's Hamiltonian}
 \eeq
 where $R_3 =
{1\/2} |+ \rangle \langle + | - {1\/2}| - \rangle \langle - |$ is
the pseudospin operator commonly used in the description of
two-level atoms\cite{Dicke}. The free Hamiltonian of the quantum
 scalar field that governs its time evolution with respect to $\tau$
is
 \beq
  H_F (\tau) = \int
d^3 k\, \omega_\k \,a^\dagger_\k\, a_\k\,
    {dt\/d \tau}\;. \label{free Hamiltonian}
 \eeq
Here $a^\dagger_\k$, $a_\k$ are the creation and annihilation
operators with momentum $\k$. The interaction between the atom and
the quantum field is assumed to be
 described by a Hamiltonian~\cite{Audretsch94}
 \beq
 H_I (\tau) = \mu\, \,R_2 (\tau)\,\phi (
x(\tau))\;, \label{interaction Hamiltonian}
 \eeq
 where $\mu$ is a
coupling constant which we assume to be small, $R_2 = {1\/2} i ( R_-
- R_+)$, and $R_+ = |+ \rangle \langle - |$, $R_- = |- \rangle
\langle +|$. The coupling is effective only on the trajectory
$x(\tau)$ of the atom.

We can now write down the Heisenberg equations of motion for the
atom and field observables. The field is always assumed to be in its
vacuum state $|0 \rangle$.  We will separately discuss the two
physical mechanisms that contribute to the rate of change of atomic
observables: the contribution of vacuum fluctuations and that of
radiation reaction. For this purpose, we can split the solution of
field $\phi$ of the Heisenberg equations  into two parts:  a free or
vacuum part $\phi^f$, which is present even in the absence of
coupling, and a source part $\phi^s$, which represents the field
generated by the interaction between the atom and the field.
Following DDC\cite{Dalibard82,Dalibard84}, we choose a symmetric
ordering between atom and field variables and consider the effects
of $\phi^f$ and $\phi^s$ separately in the Heisenberg equations of
an arbitrary atomic observable G. Then, we obtain the individual
contributions of vacuum fluctuations and radiation reaction to the
rate of change of G. Since we are interested in the spontaneous
excitation of the atom,  we will concentrate on the mean atomic
excitation energy $\langle H_A(\tau) \rangle$. The contributions of
vacuum fluctuations(vf) and radiation reaction(rr) to the rate of
change of $\langle H_A \rangle$ can be written as ( cf.
Ref.\cite{Dalibard82,Dalibard84,Audretsch94} )
 \bea
 \left\langle {d H_A (\tau) \/ d\tau}
\right\rangle_{vf} &=&
    2 i\, \mu^2  \int_{\tau_0}^\tau d \tau' \, C^F(x(\tau),x(\tau'))
    {d\/ d \tau} \chi^A(\tau,\tau')\;, \label{general form of vf}\\
\left\langle {d H_A (\tau) \/ d\tau} \right\rangle_{rr} &=&
    2 i\, \mu^2
    \int_{\tau_0}^\tau d \tau' \, \chi^F(x(\tau),x(\tau')) {d\/
d \tau}
    C^A(\tau,\tau')\;,
    \label{general form of rr}
 \eea
with $| \rangle = |a,0 \rangle$ representing the atom in the state
$|a\rangle$ and the field in the  vacuum state $|0 \rangle$. Here
the statistical functions of the atom, $C^{A}(\tau,\tau')$ and
$\chi^A(\tau,\tau')$, are defined as \bea C^{A}(\tau,\tau') &=&
{1\/2} \langle a| \{ R_2^f (\tau), R_2^f (\tau')\}
    | a \rangle\;,\label{general form of Ca} \\
\chi^A(\tau,\tau') &=& {1\/2} \langle a| [ R_2^f (\tau), R_2^f
(\tau')]
    | a \rangle \;,\label{general form of Xa}
 \eea
 and those of the field are as
 \bea
 C^{F}(x(\tau),x(\tau')) &=& {1\/2}{\langle} 0| \{ \phi^f
(x(\tau)), \phi^f(x(\tau')) \} | 0 \rangle\;, \label{general form of
Cf}\\
\chi^F(x(\tau),x(\tau')) &=& {1\/2}{\langle} 0| [
\phi^f(x(\tau)),\phi^f (x(\tau'))] | 0 \rangle\;. \label{general
form of Xf}
 \eea
$C^A$ is called the symmetric correlation function of the atom in
the state $|a\rangle$, $\chi^A$ its linear susceptibility.  $C^F$
and $\chi^F$ are the Hadamard function and Pauli-Jordan or
 Schwinger function of the field respectively.
The explicit forms of the statistical functions of the atom are
given by \bea C^{A}(\tau,\tau')&=&{1\/2} \sum_b|\langle a | R_2^f
(0) | b
    \rangle |^2 \left( e^{i \omega_{ab}(\tau - \tau')} + e^{-i
\omega_{ab}
    (\tau - \tau')} \right)\;, \label{explicit form of Ca}\\
\chi^A(\tau,\tau') & =& {1\/2}\sum_b |\langle a | R_2^f (0) | b
\rangle |^2
    \left(e^{i \omega_{ab}(\tau - \tau')} - e^{-i \omega_{ab}(\tau -
\tau')}
    \right)\;, \label{explicit form of Xa}\eea
where $\omega_{ab}= \omega_a-\omega_b$ and the sum runs over a
complete set of atomic states.

\section{Spontaneous excitation of static atoms outside a black hole.}
In the exterior region of the Schwarzschild black hole, a complete
set of normalized basis functions for the massless scalar field that
satisfy the Klein-Gordon equation  is given by
\begin{eqnarray}
 \overrightarrow{u}_{\omega
lm}=(4\pi\omega)^{-\frac{1}{2}}e^{-i\omega
t}\overrightarrow{R}_l(\omega|r)Y_{lm}(\theta,\varphi)\;,
\end{eqnarray}
\begin{eqnarray}
\overleftarrow{u}_{\omega lm}=(4\pi\omega)^{-\frac{1}{2}}e^{-i\omega
t}\overleftarrow{R}_l(\omega|r)Y_{lm}(\theta,\varphi)\;,
\end{eqnarray}
where $Y_{lm}(\theta,\varphi)$ are the spherical harmonics and the
radial functions have the following asymptotic forms~\cite{Dewitt75}
\begin{equation} \label{asymp1}
\overrightarrow{R}_l(\omega|r)\sim\left\{
                    \begin{aligned}
                 &r^{-1}e^{i\omega
r_\ast}+\overrightarrow{A}_l(\omega)r^{-1}e^{-i\omega r_\ast},\;\;r
\rightarrow 2M\;,\cr
                  &
                  {B}_l(\omega)r^{-1}e^{i\omega r_\ast},\;\;\quad\quad \quad\quad \;\;\;\;r
\rightarrow\infty\;,\cr
                          \end{aligned} \right.
\end{equation}
\begin{equation} \label{asymp2}
\overleftarrow{R}_l(\omega|r)\sim\left\{
                      \begin{aligned}
&{B}_l(\omega)r^{-1}e^{-i\omega r_\ast},\;\;\quad\quad \quad\quad
\;\;\;\;r \rightarrow2M\;,\cr &r^{-1}e^{-i\omega
r_\ast}+\overleftarrow{A}_l(\omega)r^{-1}e^{i\omega r_\ast},\;\;r
\rightarrow \infty\;,
                       \end{aligned} \right.
\end{equation}
with
\begin{eqnarray} r_\ast=r+2M\ln\bigg(\frac{r}{2M}-1\bigg)\;,
\end{eqnarray}
being the Regge-Wheeler tortoise coordinate. The physical
interpretation of these modes is that $\overrightarrow u$
represents modes emerging from the past horizon and the
$\overleftarrow u$ denotes those coming in from infinity. With the
basics of the scalar field modes given above, we now apply the
formalism outlined in the preceding section to examine the
spontaneous excitation of the static atoms in three vacuum states
of the quantum scalar fields respectively.

\paragraph{Boulware vacuum.} The Boulware vacuum is defined by requiring normal modes to be
positive frequency with respect to the Killing vector $\partial/
\partial t$. One can  show that the Wightman function for
massless scalar fields in this vacuum state is given
by~\cite{Fulling77,Candelas80}
 \bea
D_B^+(x,x')\,=\frac{1}{4\pi}\sum_{lm}|Y_{lm}(\theta,\varphi)|^2\,
    \int_{0}^{+\infty}\frac{d\omega}{\omega}\,
    e^{-i\omega\Delta t}\biggl[\,|\overrightarrow{R}_l(\omega|\,r)|^2
    +|\overleftarrow{R}_l(\omega|\,r)|^2\biggr]\;,
\eea
 and the corresponding Hadamard function and Pauli-Jordan or
 Schwinger function of the field are respectively
 \bea
C^F(x\,(\tau),x\,(\tau')\,)&=&\frac{1}{8\pi}\,\sum_{lm}\,|Y_{lm}(\theta,\varphi)|^2
   \int_0^{+\infty}\frac{d\omega}{\omega}\,
   \biggl(e^{\frac{i\omega\Delta\tau}{\sqrt{1-2M/r}}}+e^{-\frac{i\omega\Delta\tau}
   {\sqrt{1-2M/r}}}\biggr)\times\nonumber\\&&\biggl[|\overrightarrow{R}_l(\omega|\,r)|^2
   +|\overleftarrow{R}_l(\omega|\,r)|^2\biggr]\;,
\eea
 and
 \bea
\chi^F(x\,(\tau),x\,(\tau')\,)&=&\frac{1}{8\pi}\,\sum_{lm}\,|Y_{lm}(\theta,\varphi)|^2
   \int_0^{+\infty}\frac{d\omega}{\omega}\,
   \biggl(e^{-\frac{i\omega\Delta\tau}{\sqrt{1-2M/r}}}-e^{\frac{i\omega\Delta\tau}{\sqrt{1-2M/r}}}
   \biggr)\times\nonumber\\&&\biggl[|\overrightarrow{R}_l(\omega|\,r)|^2
   +|\overleftarrow{R}_l(\omega|\,r)|^2\biggr]\;,
\eea
 where use has been made of
\begin{equation}
 \Delta\tau=\Delta\,t \,\sqrt{1-\frac{2M}{r}}\;.
 \end{equation}
Substituting the above results into Eqs.~(\ref{general form of vf})
and (\ref{general form of rr}),  extending the integration range for
$\tau$ to infinity for sufficiently long times $\tau-\tau_0$, and
performing the double integration, we obtain
 the contribution of the vacuum fluctuations to the rate of change
of the mean atomic energy for an atom held static at a distance $r$
from the black hole
\bea
\biggl\langle\frac{dH_A(\tau)}{d\tau}\biggr\rangle_{vf}&=&
   -\,\frac{\mu^2}{4\pi}\,\biggl[\;\sum_{\omega_a>\omega_b}
   \, \omega_{ab}^2\,|\langle
   a|R_2^f(0)|b\rangle|^2\,P\,(\,\omega_{ab}\,,r)\nonumber\\&&\;\quad\quad-\sum_{\omega_a<\omega_b}
   \, \omega_{ab}^2\,|\langle a|R_2^f(0)|b\rangle|^2 P\,(-\,\omega_{ab}\,,r)\biggr]\;,
\eea
 and that of radiation reaction
\bea
\biggl\langle\frac{dH_A(\tau)}{d\tau}\biggr\rangle_{rr}&=&
   -\,\frac{\mu^2}{4\pi}\,\biggl[\;\sum_{\omega_a>\omega_b}
   \, \omega_{ab}^2\,|\langle
   a|R_2^f(0)|b\rangle|^2\,P\,(\,\omega_{ab}\,,r)\nonumber\\&&\;\quad\quad+\sum_{\omega_a<\omega_b}
   \, \omega_{ab}^2\,|\langle a|R_2^f(0)|b\rangle|^2
   P\,(-\,\omega_{ab}\,,r)\biggr]\;.
\eea
 Here we have defined
 \bea
P(\omega_{ab},r)= \overrightarrow{P}(\omega_{ab},r) +
\overleftarrow{P}(\omega_{ab},r)\;,
 \eea
 \bea
 \label{rightP}
 \overrightarrow{P}(\omega_{ab},r)&=&{\pi \over
   \omega_{ab}^2}\sum_{lm}|Y_{lm}(\theta,\phi)|^2\,\biggl|\overrightarrow
   R_l\biggl(\omega_{ab}r\sqrt{1-\frac{2M}{r}}\biggr)\;\biggr|^2\nonumber\\
   &=&\,\frac{1}{\omega_{ab}^2}\sum_{l=0}^{\infty}\frac{2l+1}{4}\;\biggl|\overrightarrow
   R_l\biggl(\omega_{ab}r\sqrt{1-\frac{2M}{r}}\biggr)\biggr|^2\;,
 \eea
 and
 \bea
 \label{leftP}
 \overleftarrow{P}(\omega_{ab},r)&=&{\pi \over
   \omega_{ab}^2}\sum_{lm}|Y_{lm}(\theta,\phi)|^2\,\biggl|\overleftarrow
   R_l\biggl(\omega_{ab}r\sqrt{1-\frac{2M}{r}}\biggr)\biggr|^2\nonumber\\
   &=&\,\frac{1}{\omega_{ab}^2}\sum_{l=0}^{\infty}\frac{2l+1}{4}\;\biggl|\overleftarrow
   R_l\biggl(\omega_{ab}r\sqrt{1-\frac{2M}{r}}\biggr)\biggr|^2\;.
 \eea
The following property of the spherical harmonics
\begin{equation}
 \sum^l_{m=-l}|\,Y_{lm}(\,\theta,\varphi\,)\,|^2= {2l+1 \over
 4\pi}\;.
 \end{equation}
has been utilized in Eqs.~(\ref{rightP}) and (\ref{leftP}). Adding
up two contributions, we obtain the total rate of change of the mean
atomic energy
 \bea
 \label{BoulwareRate}
\biggl\langle\frac{dH_A(\tau)}{d\tau}\biggr\rangle_{tot}\,=
   -\,\frac{\mu^2}{2\pi}\sum_{\omega_a>\omega_b}
   \, \omega_{ab}^2\,|\langle
   a|R_2^f(0)|b\rangle|^2\,P\,(\,\omega_{ab}\,,r)\;.
 \eea
It follows that for an static atom in the ground state
$(\omega_a<\omega_b)$, the contribution of the vacuum fluctuations
and that of radiation exactly cancel, since each term in
$\left\langle {d H_A (\tau) \/ d\tau} \right\rangle_{vf}$ is
canceled exactly by the corresponding term in $\left\langle {d H_A
(\tau) \/ d\tau} \right\rangle_{rr}$. Therefore, although both
contributions to the rate of change of the mean atomic energy  is
modified by the presence of the factor $P(\omega_{ab}, r )$ as
compared to the Minkowski vacuum case~\cite{Audretsch94}, the
balance between them remains and the static ground state atom in the
Boulware vacuum is still stable. It should be pointed out, however,
that the spontaneous emission rate of a static atom outside a
Schwarzschild black hole in the Boulware vacuum is different from
that of an inertial atom in the Minkowski vacuum in an unbounded
flat space because of the presence of  the factor $P(\omega_{ab}, r
)$ in Eq.~(\ref{BoulwareRate}). In this sense, the Boulware vacuum
is not equivalent to the usual Minkowski vacuum. However, a
comparison of Eq.~(\ref{BoulwareRate}) with Eq.~(23) in
Ref.~\cite{H. Yu} , which gives the rate of change of the mean
atomic energy for an inertial atom in a flat space with a reflecting
boundary, shows that the two rates are quite similar, and the
appearance of $P(\omega_{ab}, r )$ in Eq.~(\ref{BoulwareRate}) can
be understood as a result of backscattering of the vacuum field
modes off the spacetime curvature of the black hole in much the same
way as the reflection of the field modes at the reflecting boundary
in a flat spacetime. In order to gain more understanding, let us now
analyze the behavior of $P(\omega_{ab}, r )$ both in the asymptotic
region and at the event horizon. Using the following asymptotic
properties of the radial functions
\begin{equation} \label{asymp3}
\sum_{l=0}^\infty\,(2l+1)\,|\overrightarrow{R}_l(\,\omega\,|r\,)\,|^2\sim\left\{
                    \begin{aligned}
                 &\frac{4\omega^2}{1-\frac{2M}{r}}\;,\;\;\;\quad\quad\quad\quad\quad\quad\quad r\rightarrow2M\;,\cr
                  &\frac{1}{r^2}
\sum_{l=0}^\infty(2l+1)\,|\,{B}_l\,(\omega)\,|^2\;,\quad\;r\rightarrow\infty
                  \;,\cr
                          \end{aligned} \right.
\end{equation}
\begin{equation} \label{asymp4}
\sum_{l=0}^\infty\,(2l+1)\,|\overleftarrow{R}_l(\,\omega\,|r\,)\,|^2\sim\left\{
                    \begin{aligned}
                 &\frac{1}{4M^2}\sum_{l=0}^\infty(2l+1)\,|\,{B}_l\,(\omega)\,|^2,\quad\;r\rightarrow2M\;,\cr
                  &4\omega^2,\;\;\;\;\quad\quad\quad\quad\quad\quad\quad\quad\quad\quad r\rightarrow\infty
                  \;,\cr
                          \end{aligned} \right.
\end{equation}
we obtain \beq \label{asymp rightP}
    \overrightarrow{P}(\,\omega_{ab},r)\sim\left\{
                    \begin{aligned}
                    &1\;,\;\;\;\;\quad\quad\quad\quad\quad\quad\quad\quad\quad\quad\quad
                    \quad\;r\rightarrow2M\;,\cr
                    &\frac{1}{4r^2\omega_{ab}^2}
   \,\sum_{l=0}^\infty\,(2l+1)\,|\,B_l\,(\,\omega_{ab})|^2\;,\;\;\; r\rightarrow\infty
                  \;,\cr
                          \end{aligned} \right.
\eeq
 \beq \label{asymp leftP}
    \overleftarrow{P}(\,\omega_{ab},r)\sim\left\{
                    \begin{aligned}
                    &\frac{1}{16M^2\omega_{ab}^2}
    \sum_{l=0}^\infty(2l+1)\,|\,{B}_l\,(\,0\,)|^2\;,\;\;\;r\rightarrow2M\;,\cr
                    &1\;,\;\;\;\;\quad\quad\quad\quad\quad\quad\quad\quad\quad
                    \quad\quad\quad\;\;r\rightarrow\infty\;,\cr
\end{aligned} \right.
\eeq
 and this leads to
 \beq \label{asymp_P}
P(\omega_{ab}, r)\sim\left\{
                    \begin{aligned}
                    &1+\frac{1}{16M^2\omega_{ab}^2}
    \sum_{l=0}^\infty(2l+1)\,|\,{B}_l\,(\,0\,)\,|^2\;,\;\;r\rightarrow2M\;,\cr
                    &1+\frac{1}{4r^2\omega_{ab}^2}
   \,\sum_{l=0}^\infty\,(2l+1)\,|\,B_l\,(\,\omega_{ab})\,|^2\;,\;\; r\rightarrow\infty
                  \;.\cr
                          \end{aligned} \right.
 \eeq
 So, when\,$r\rightarrow\infty$, we have
 \bea
 \biggl\langle\frac{dH_A(\tau)}{d\tau}\biggr\rangle_{tot}\approx
   -\,\frac{\mu^2}{2\pi}\sum_{\omega_a>\omega_b}\, \omega_{ab}^2\,|\langle
   a|R_2^f(0)|b\rangle|^2\,\biggl[1+\frac{1}{4r^2\omega_{ab}^2}
   \,\sum_{l=0}^\infty\,(2l+1)\,|\,B_l\,(\,\omega_{ab})|^2\biggr]\;,
\eea
 and when $r\rightarrow2M$,
 \bea
 \label{BRateEH}
 \biggl\langle\frac{dH_A(\tau)}{d\tau}\biggr\rangle_{tot}\approx
   -\,\frac{\mu^2}{2\pi}\sum_{\omega_a>\omega_b}\, \omega_{ab}^2\,|\langle
   a|R_2^f(0)|b\rangle|^2\,\biggl[1+\frac{1}{16M^2\omega_{ab}^2}
    \sum_{l=0}^\infty(2l+1)\,|\,{B}_l\,(\,0\,)\,|^2\;\biggr]\;.
\eea
 These asymptotic forms tell us that the rate of change of the
mean atomic energy for a static atom outside a Schwarzschild black
hole interacting with massless scalar fields in the Boulware vacuum
gets enhanced as compared to the case of an inertial atom in the
Minkowski vacuum in an unbounded flat space, and it reduces to the
result of an inertial atom in the Minkowski vacuum at infinity and
behaves normally at the event horizon. The normal behavior of the
rate of change of the mean atomic energy near the horizon is in
sharp contrast to the response rate of an Unruh
detector~\cite{Candelas80}.

\paragraph{Unruh vacuum.} For the Unruh vacuum, the Wightman function for
the massless scalar fields is given by~\cite{Fulling77,Candelas80}
 \bea
D_U^+(x,x')\,&=&\frac{1}{4\pi}\sum_{lm}|Y_{lm}(\theta,\varphi)|^2\,
    \int_{-\infty}^{+\infty}\frac{d\omega}{\omega}\times\nonumber\\&&
    \biggl[\,\frac{e^{-i\omega\Delta t}}{1-e^{-2\pi\,\omega/\kappa}}\,
    |\overrightarrow{R}_l(\omega|\,r)|^2
    +\theta(\omega)\,e^{-i\omega\Delta
    t}|\overleftarrow{R}_l(\omega|\,r)|^2\biggr]\;,
\eea
 where $\kappa=1/4M$ is the surface gravity of the black hole.
Then the statistical functions of the scalar field readily follow

 \bea
C^F(x\,(\tau),x\,(\tau')\,)&=&\frac{1}{8\pi}\,\sum_{lm}\,|Y_{lm}(\theta,\varphi)|^2
   \int_{-\infty}^{+\infty}\frac{d\omega}{\omega}\,
   \biggl(e^{\frac{i\omega\Delta\tau}{\sqrt{1-2M/r}}}+e^{-\frac{i\omega\Delta\tau}
   {\sqrt{1-2M/r}}}\biggr)\times\nonumber\\&&\biggl(\frac{|\overrightarrow{R}_l(\omega|\,r)|^2}
   {{1-e^{-2\pi\,\omega/\kappa}}}
   +\theta(\omega)|\overleftarrow{R}_l(\omega|\,r)|^2\biggr)\;,
\eea
 \bea
\chi^F(x\,(\tau),x\,(\tau')\,)&=&\frac{1}{8\pi}\sum_{lm}\,|Y_{lm}(\theta,\varphi)|^2\,
   \int_{-\infty}^{+\infty}\frac{d\omega}{\omega}\,\biggl(e^{-\frac{i\omega\Delta\tau}
   {\sqrt{1-2M/r}}}-e^{\frac{i\omega\Delta\tau}{\sqrt{1-2M/r}}}\biggr)\,\times
   \nonumber\\&&\biggl[\frac{|\overrightarrow{R}_l(\omega|\,r)|^2}{{1-e^{-2\pi\,\omega/\kappa}}}+
   \theta(\omega)|\overleftarrow{R}_l(\omega|\,r)|^2\biggr]\;.
\eea
 Similarly, we can compute the contributions of vacuum fluctuations
 and radiation reaction to the rate of change of the mean atomic energy to get
   \bea
\biggl\langle\frac{dH_A(\tau)}{d\tau}\biggr\rangle_{vf}&=&-\,\frac{\mu^2}{4\pi}\,
   \biggl\{\sum_{\omega_a>\omega_b}\,\omega_{ab}^2|\langle a|R_2^f(0)|b\rangle|^2\,
   \biggl[\,\biggl(1+\frac{1}{e^{(2\pi\,\omega_{ab})/\kappa_r}-1}\biggr)
   \overrightarrow{P}(\omega_{ab},r)\nonumber\\&&\;\;\;\;\quad\quad\quad\quad+\,
   \frac{\overrightarrow{P}(-\,\omega_{ab},r)}{e^{(2\pi\,\omega_{ab})/\kappa_r}-1}\,+
   \overleftarrow{P}(\omega_{ab},r)\,\biggr]\nonumber\\&&\;\;\quad\quad-
   \sum_{\omega_a<\omega_b}\omega_{ab}^2|\langle a|R_2^f(0)|b\rangle|^2
   \biggl[\biggl(\,1+\frac{1}{e^{(2\pi\,|\,\omega_{ab}|)/\kappa_r}-1}\biggr)
   \overrightarrow{P}(-\,\omega_{ab},r)\nonumber\\&&\,\;\;\;\;\quad\quad\quad\quad+\,
   \frac{\overrightarrow{P}(\omega_{ab},r)}{e^{{(2\pi\,|\,\omega_{ab}|)/\kappa_r}-1}}
   +\overleftarrow{P}(-\,\omega_{ab},r)\,\biggl]\biggr\}\;,
\eea
 and
 \bea
\biggl\langle\frac{dH_A(\tau)}{d\tau}\biggr\rangle_{rr}&=&-\,\frac{\mu^2}{4\pi}\,
   \biggl\{\sum_{\omega_a>\omega_b}\omega_{ab}^2|\langle a|R_2^f(0)|b\rangle|^2\,
   \biggl[\,\biggl(1+\frac{1}{e^{(2\pi\,\omega_{ab})/\kappa_r}-1}\biggr)
   \overrightarrow{P}(\omega_{ab},r)\nonumber\\&&\;\;\;\;\quad\quad\quad\quad-\,
   \frac{\overrightarrow{P}(-\,\omega_{ab},r)}{e^{(2\pi\,\omega_{ab})/\kappa_r}-1}+
   \overleftarrow{P}(\omega_{ab},r)\,\biggr]\nonumber\\&&\;\;\quad\quad+\sum_{\omega_a<\omega_b}
   \omega_{ab}^2|\langle a|R_2^f(0)|b\rangle|^2
   \biggl[\,\biggl(\,1+\frac{1}{e^{(2\pi\,|\,\omega_{ab}|)/\kappa_r}-1}
   \biggr)\overrightarrow{P}(-\,\omega_{ab},r)\nonumber\\&&\,\;\;\;\;\quad\quad\quad\quad-\,
   \frac{\overrightarrow{P}(\omega_{ab},r)}{e^{(2\pi\,|\,\omega_{ab}|)/\kappa_r}-1}
   +\overleftarrow{P}(-\,\omega_{ab},r)\,\biggl]\,\biggr\}\;,
\eea
 where we have defined
 \begin{equation}
\kappa_r=\frac{\kappa}{\sqrt{1-\frac{2M}{r}}}\;.
 \end{equation}
 From the above results, one can see that both contributions are
 altered due to the appearance of thermal terms, as compared to the case of the Boulware vacuum. If we add
 up two contributions, we find the total rate
 \bea
\biggl\langle\frac{dH_A(\tau)}{d\tau}\biggr\rangle_{tot}&=&-\,\frac{\mu^2}{2\pi}\,
   \biggl\{\sum_{\omega_a>\omega_b}\omega_{ab}^2|\langle a|R_2^f(0)|b\rangle|^2\,\biggl[\,
   \biggl(1+\frac{1}{e^{(2\pi\,\omega_{ab})/\kappa_r}-1}\biggr)\,
   \overrightarrow{P}(\omega_{ab})+\,\overleftarrow{P}(\omega_{ab})\,\biggr]\nonumber\\&&
   \;\;\quad\quad-\sum_{\omega_a<\omega_b}\omega_{ab}^2|\langle\,a|R_2^f(0)|b\rangle|^2\,\,
   \frac{\overrightarrow{P}(\omega_{ab})}{e^{(2\pi\,|\,\omega_{ab}|)/\kappa_r}-1}\,
   \,\biggr\}\;.
\eea
 This reveals that  the delicate balance no longer exists between the vacuum fluctuations and radiation
reaction that ensures the stability of ground state atoms held
static at a radial distance $r$ from the black hole in the Boulware
vacuum. There is a positive contribution from the second term (
$\omega_{a}< \omega_{b}$ term), therefore transitions of
ground-state atoms to excited states could spontaneously occur in
the Unruh vacuum outside the black hole.

When the atom is held close to the event horizon, i.e.,
when\;$r\rightarrow2M$, the total rate becomes
  \bea
  \label{URateEH}
\biggl\langle\frac{dH_A(\tau)}{d\tau}\biggr\rangle_{tot}&\approx&-\,\frac{\mu^2}{2\pi}\,
   \biggl\{\sum_{\omega_a>\omega_b}|\langle a|R_2^f(0)|b\rangle|^2\omega_{ab}^2\,\times\nonumber\\&&
   \quad\quad\quad\biggl[\,\biggl(1+\frac{1}{16\,M^2\omega_{ab}^2}\,\sum_{l=0}^\infty\,(2l+1)\,|\,B_l\,(\,0\,)|^2
   \biggr)+ \frac{1}{e^{(2\pi\,\omega_{ab})/\kappa_r}-1}
   \biggr]\nonumber\\&&
   \quad\quad\;-\sum_{\omega_a<\omega_b}|\langle\,a|R_2^f(0)|b\rangle|^2\,
   \omega_{ab}^2\,\frac{1}{e^{(2\pi\,|\,\omega_{ab}|)/\kappa_r}-1}
   \biggr\}\;.
\eea
 In comparison to Eq.~(\ref{BRateEH}), the corresponding result in the Boulware vacuum case, one
 sees the appearance of thermal terms which may be considered as resulting from the
 contribution of thermal radiation emanating from the black hole at a temperature
 \bea
T={\kappa_r\over
2\pi}=\frac{\kappa}{2\pi}\frac{1}{\sqrt{1-\frac{2M}{r}}}
  =(g_{00})^{-1/2}\,T_{H}\;,
\eea
 where $T_H=\kappa/2\pi$ is the usual Hawking temperature of the
black hole. Actually, this is  the well-known Tolman
relation~\cite{Tolman} which gives the proper temperature as
measured by a local observer. Notice that  $T$, being always larger
than the Hawking temperature, and  reducing to it only at infinity,
 however diverges as the event horizon is approached. This can be
 understood as a result of that the atom must be in acceleration relative to the local free-falling frame to maintain at
a fixed distance from the black hole, and this acceleration, which
blows up at the horizon, gives rise to additional thermal effect.

If the atom is far away from the black hole in the asymptotic
region, that is, when\,$r\rightarrow\infty$, one then finds
   \bea
\biggl\langle\frac{dH_A(\tau)}{d\tau}\biggr\rangle_{tot}&\approx&-\,{\mu^2\over2\pi}\,
   \biggl\{\sum_{\omega_a>\omega_b}|\langle a|R_2^f(0)|b\rangle|^2\omega_{ab}^2\,\biggl[\,1+f(\omega_{ab},
r) + \frac{f(\omega_{ab}, r)}{e^{(2\pi\,\omega_{ab})/\kappa_r}-1}
   \,\biggr]
   \nonumber\\&&\;\quad\quad-\sum_{\omega_a<\omega_b}|\langle\,a|R_2^f(0)|b\rangle|^2\,
   \omega_{ab}^2\,\frac{f(\omega_{ab},
r)}{e^{(2\pi\,|\,\omega_{ab}|)/\kappa_r}-1}\,
   \biggr\}\;,
\eea
 where
 \bea
f(\omega_{ab},
r)=\frac{1}{4\,r^2\omega_{ab}^2}\,\sum_{l=0}^\infty\,(2l+1)\,|\,B_l\,(\,\omega_{ab})|^2\;.
 \eea
The appearance of $f(\omega_{ab}, r)$ in the thermal terms now can
be envisaged as a result of backscattering of outgoing thermal flux
from the event horizon off the spacetime curvature. The
backscattering results in the depletion of part of the outgoing
flux. The influence of the thermal flux becomes weaker as the atom
is placed farther away.

\paragraph{Hartle-Hawking vacuum.}  Let us now turn briefly to the case of the Hartle-Hawking vacuum.
The Wightman function for the massless scalar fields becomes now
~\cite{Fulling77,Candelas80}
\begin{eqnarray}
D_H^+(x,x')\,=\frac{1}{4\pi}\sum_{lm}\,|Y_{lm}(\theta,\varphi)|^2\
    \int_{-\infty}^{+\infty}\frac{d\omega}{\omega}\,\biggl[\,\frac{e^{-{i\omega\Delta t}
    }}{1-e^{-2\pi\,\omega/\kappa}}\,|\overrightarrow{R}_l(\omega|r)|^2
    +\frac{e^{-{i\omega\Delta t}}}{1-e^{-2\pi\,\omega/\kappa}}\,|\overleftarrow{R}_l(\omega|r)|^2\biggr]\;,\nonumber\\
\end{eqnarray}
which leads to the statistical functions of the scalar field in the
Hartle-Hawking vacuum as follows
\begin{eqnarray}
C^F(x\,(\tau),x\,(\tau')\,)&=&\frac{1}{8\pi}\sum_{lm}\,|Y_{lm}(\theta,\varphi)|^2\,
   \int_{-\infty}^{+\infty}\frac{d\omega}{\omega}\,\biggl(e^{\frac{i\omega\Delta\tau}
   {\sqrt{1-{2M}/{r}}}}+e^{-\frac{i\omega\Delta\tau}{\sqrt{1-{2M}/{r}}}}\biggr)\,
   \nonumber\\&&\times\,\biggl(\frac{|\overrightarrow{R}_l(\omega|r)|^2}{{1-e^{-{2\pi\,\omega}/{\kappa}}}}
   +\frac{|\overleftarrow{R}_l(\omega|r)|^2}{{e^{{2\pi\,\omega}/{\kappa}}-1}}\biggr)\;,
\end{eqnarray}
and
\begin{eqnarray}
\chi^F(x\,(\tau),x\,(\tau')\,)&=&\frac{1}{8\pi}\sum_{lm}\,|Y_{lm}(\theta,\varphi)|^2\,
   \int_{-\infty}^{+\infty}\frac{d\omega}{\omega}\,\biggl(e^{\frac{i\omega\Delta\tau}
   {\sqrt{1-{2M}/{r}}}}-e^{-\frac{i\omega\Delta\tau}{\sqrt{1-{2M}/{r}}}}\biggr)\,
   \nonumber\\&&\times\,\biggl(\frac{|\overleftarrow{R}_l(\omega|r)|^2}{{e^{{2\pi\,\omega}/{\kappa}}-1}}
   -\frac{|\overrightarrow{R}_l(\omega|r)|^2}{{1-e^{-{2\pi\,\omega}/{\kappa}}}}\biggr)\;.
\end{eqnarray}
By using the above results and Eqs.~(\ref{general form of vf})and
~(\ref{general form of rr}),
 the contribution of the vacuum fluctuations to the rate of change
of the mean atomic energy can be found for an atom held static at a
distance $r$ from the black hole
\begin{eqnarray}
\biggl\langle\frac{dH_A(\tau)}{d\tau}\biggr\rangle_{vf}&=&-\,\frac{\mu^2}{4\pi}\,\biggl\{\sum_{\omega_a>\omega_b}
   \, \omega_{ab}^2\,|\langle
a|R_2^f(0)|b\rangle|^2\,\biggl[\;\;\frac{P(-\,\omega_{ab},r)}{e^{(2\pi\,\omega_{ab})/{\kappa_r}}-1}
   \nonumber\\
   &&\quad\quad\quad\quad\quad\;\;\;\;+\biggl(1+\frac{1}{e^{(2\pi\,\omega_{ab})/{\kappa_r}}-1}\biggr)
   P(\omega_{ab},r)\,\biggr]
   \nonumber\\&&-\sum_{\omega_a<\omega_b} \, \omega_{ab}^2\,|\langle
   a|R_2^f(0)|b\rangle|^2\,\biggl[\;\;
   \frac{P\,(\omega_{ab}, r)}{e^{({2\pi\,|\,\omega_{ab}|})/{\kappa_r}}-1}
    \nonumber\\
   &&\quad\quad\quad\quad\quad\;\;\;\;+\,\biggl(\,1+
   \frac{1}{e^{({2\pi\,|\,\omega_{ab}|})/{\kappa_r}}-1}\biggr)\,P\,(-\,\omega_{ab}, r)\,\biggl]\,\biggr\}
   \;, \nonumber\\
\end{eqnarray}
and that of radiation reaction
\begin{eqnarray}
\biggl\langle\frac{dH_A(\tau)}{d\tau}\biggr\rangle_{rr}&=&-\,\frac{\mu^2}{4\pi}\,\biggl\{\sum_{\omega_a>\omega_b}
   \, \omega_{ab}^2\,|\langle a|R_2^f(0)|b\rangle|^2\,\biggl[\;\;-\frac{P(-\,\omega_{ab},r)}{e^{(2\pi\,\omega_{ab})/{\kappa_r}}-1}
   \nonumber\\
   &&\quad\quad\quad\quad\quad\;\;\;\;+\biggl(1+\frac{1}{e^{(2\pi\,\omega_{ab})/{\kappa_r}}-1}\biggr)
   P(\omega_{ab},r)\,\biggr]
   \nonumber\\&&-\sum_{\omega_a<\omega_b}\, \omega_{ab}^2\,|\langle
   a|R_2^f(0)|b\rangle|^2\,\biggl[\;\;
   \frac{P\,(\omega_{ab}, r)}{e^{({2\pi\,|\,\omega_{ab}|})/{\kappa_r}}-1}
    \nonumber\\
   &&\quad\quad\quad\quad\quad\;\;\;\;-\,\biggl(\,1+
   \frac{1}{e^{({2\pi\,|\,\omega_{ab}|})/{\kappa_r}}-1}\biggr)\,P\,(-\,\omega_{ab}, r)\,\biggl]\,\biggr\}
   \;.\nonumber\\
\end{eqnarray}
  Consequently, the total rate of change
of the mean atomic energy follows
\begin{eqnarray}
\biggl\langle\frac{dH_A(\tau)}{d\tau}\biggr\rangle_{tot}&=&-\,\frac{\mu^2}{2\pi}\,\biggl[\sum_{\omega_a>\omega_b}
   \, \omega_{ab}^2\,|\langle a|R_2^f(0)|b\rangle|^2\, P\,(\omega_{ab},r )\;\biggl(1+\frac{1}{e^{(2\pi\,\omega_{ab})/{\kappa_r}}-1}\biggr)\,
  \nonumber\\&&-\sum_{\omega_a<\omega_b}\, \omega_{ab}^2\,|\langle
  a|R_2^f(0)|b\rangle|^2\,P(\omega_{ab}, r)\,
   \frac{1}{e^{({2\pi\,|\,\omega_{ab}|})/{\kappa_r}}-1}\,\biggr]\;.
\end{eqnarray}
Once again, with the existence of $\omega_a<\omega_b$ term, for
static atoms in the Hartle-Hawking vacuum, transitions from ground
state to the excited states can occur spontaneously in the exterior
region of the black hole. In the spatial asymptotic region, the
total rate can be written as
 \bea
\biggl\langle\frac{dH_A(\tau)}{d\tau}\biggr\rangle_{tot}\,&\approx&-\,\frac{\mu^2}{2\pi}\,\biggl[\sum_{\omega_a>\omega_b}
    \omega_{ab}^2\,|\langle
    a|R_2^f(0)|b\rangle|^2\times\nonumber\\&&\quad\;\quad\;\quad\;\quad\;
    \biggl(1+\frac{1}{4r^2\omega_{ab}^2}\,\sum_{l=0}^\infty\,(2l+1)\,|\,B_l\,(\,\omega_{ab})\,|^2\biggr)\,\biggl(\,1+\frac{1}{e^{(2\pi\omega_{ab})/\kappa}-1}
    \biggr)
    \nonumber\\&&\quad\;\quad-\sum_{\omega_a<\omega_b}\omega_{ab}^2\,|\langle a|R_2^f(0)|b\rangle|^2\,
    \biggl(1+\frac{1}{4r^2\omega_{ab}^2}\,
    \sum_{l=0}^\infty\,(2l+1)\,|\,B_l\,(\,\omega_{ab})\,|^2\biggr)\,\times\nonumber\\&&\quad\;\quad\;\quad\;\quad\;\frac{1}{e^{(2\pi|\omega_{ab}|)/\kappa}-1}
    \,\biggr]\;.
\eea
  For an atom at spatial infinity ($r\rightarrow \infty$),
$P(\omega_{ab}, r)\rightarrow 1$, and the temperature as perceived
by the atom, $T$, approaches $T_H$,  and the total rate of change of
the mean atomic energy becomes what one would get if the atom in
immersed in a thermal bath at the temperature $T_H$. Therefore, an
static atom in the spatial asymptotic region outside the black hole
would spontaneously excite as if in a thermal bath of radiation at
the Hawking temperature. This is consistent with our  understanding
gained from the calculations of expectation values of
energy-momentum tensor~\cite{Candelas80} that the Hartle-Hawking
vacuum is not a state empty at infinity but corresponds instead to a
thermal distribution of (Minkowski-type) quanta at the Hawking
temperature, and therefore it describes a black hole in equilibrium
with an infinite sea of black-body radiation. On the other hand,
when the atom is held near the event horizon, i.e., when
$r\rightarrow 2M$, we have
 \bea
\biggl\langle\frac{dH_A(\tau)}{d\tau}\biggr\rangle_{tot}\,&\approx&-\,\frac{\mu^2}{2\pi}\,\biggl[\sum_{\omega_a>\omega_b}
    \omega_{ab}^2|\langle a|R_2^f(0)|b\rangle|^2\times\nonumber\\&&\quad\;\quad\;\quad\;\quad\;
    \biggl(\,1+\frac{1}{16M^2\omega_{ab}^2}\sum_{l=0}^\infty(2l+1)\,|\,{B}_l\,(\,0\,)\,
    |^2\biggr)\,\biggl(\,1+\frac{1}{e^{(2\pi\omega_{ab})/\kappa_r}-1}\biggr)\,\nonumber\\&&\quad\;\quad-\sum_{\omega_a<\omega_b}\omega_{ab}^2|\langle a|R_2^f(0)|b\rangle|^2
    \,\biggl(\,1+\frac{1}{16M^2\omega_{ab}^2}\sum_{l=0}^\infty(2l+1)\,
    |\,{B}_l\,(\,0\,)\,|^2\biggr)\,\times\nonumber\\&&\quad\;\quad\;\quad\;\quad\;
    \frac{1}{e^{(2\pi|\omega_{ab}|)/\kappa_r}-1}\,\biggr]\;.
\eea
 Here one can see that close to the horizon, in addition to the
 contribution that can be account for by the outgoing thermal
 radiation emanating from the horizon (refer to Eq.~(\ref{URateEH})), there is another
 contribution (the thermal term multiplied by the term containing ${B}_l$)
 that can be regarded as resulting from the incoming radiation from
 the sea of thermal radiation at infinity and this incoming
 radiation is however deflected by the spacetime geometry. Notice
 that the difference between the rate of change of the mean atomic
 energy in the Unruh vacuum and that in the Hartle-Hawking vacuum is not a simple factor 2 as in the 1+1 dimensional case~\cite{YuZhou07}.
 The reason is that in the four dimensional case, there are
 backscatterings by the spacetime curvature so that
 the outgoing thermal radiation from the event horizon can not travel through the spacetime
 unaffected and so does the incoming thermal radiation from
 infinity.

\section{Summary}

Using the DDC formalism, we have studied the spontaneous excitation
of a two-level atom held static outside a Schwarzschild black hole
and in interaction with a massless scalar field in the Boulware,
Unruh and Hartle-Hawking vacuum respectively, and calculated the
contributions of the vacuum fluctuations and radiation reaction to
the rate of change of the mean atomic energy.

In the Boulware vacuum case, spontaneous excitation can not occur so
that the ground state atoms are stable. However, the spontaneous
emission rate for excited atoms in the Boulware vacuum is not the
same as that in the usual Minkowski vacuum, but very similar to that
in the vacuum in a flat spacetime with a reflecting boundary. A
noteworthy feature here is that the rate of change of the mean
atomic energy is well-behaved at the event horizon, in sharp
contrast to the response rate of an Unruh
detector~\cite{Candelas80}.

Both for the Unruh vacuum and the Hartle-Hawking vacuum, our results
show that an atom held static at a radial distance $r$ from a
Schwarzschild black hole would spontaneously excite.  For the Unruh
vacuum, it spontaneously excites as if there were  an outgoing
thermal flux of radiation (backscattered by the spacetime geometry
though) at a temperature characterized by the Tolman relation. For
the Hartle-Hawking vacuum, the spontaneous excitation occurs as if
the atom were in a thermal bath of radiation at a proper temperature
which reduces to the  Hawking temperature  in the spatial asymptotic
region, except for a frequency response distortion caused by the
backscattering of the field modes off the spacetime curvature.

\begin{acknowledgments}
This work was supported in part  by the National Natural Science
Foundation of China  under Grant No.10575035 and the Program for New
Century Excellent Talents in University (NCET, No. 04-0784).

\end{acknowledgments}


\begin{thebibliography}{90}


\bibitem{Welton48}T. A. Welton, Phys. Rev. {\bf 74}, 1157(1948).
\bibitem{CPP83}G. Compagno, R. Passante and F. Persico, Phys. Lett. A {\bf
98},253(1983).
\bibitem{Ackerhalt73}J. R. Ackerhalt, P. L. Knight and J. H. Eberly, Phys. Rev.
Lett. {\bf 30}, 456(1973).
\bibitem{Milonni88}P. W. Milonni, Phys. Scr. {\bf T21}, 102(1988); P. W.
Milonni and W. A. Smith, Phys. Rev. A {\bf 11}, 814(1975).
\bibitem{Dalibard82} J. Dalibard, J. Dupont-Roc and C. Cohen-Tannoudji, J.
Phys. (France){\bf43}, 1617(1982).
\bibitem{Dalibard84} J. Dalibard, J. Dupont-Roc and C. Cohen-Tannoudji, J.
Phys. (France){\bf45}, 637(1984).

\bibitem{Audretsch94}J. Audretsch and R. M\"uller, Phys. Rev. A {\bf 50},
1755(1994).
\bibitem{H. Yu} H. Yu and S. Lu, Phys. Rev. D {\bf72},
064022(2005); {\it ibid}, D{\bf73}, 109901(2006).
\bibitem{ZYL06} Z. Zhu, H. Yu and S. Lu, Phys. Rev. D {\bf73},
107501(2006).
\bibitem{YuZ06} H. Yu and Z. Zhu, Phys. Rev. D {\bf74},
044032 (2006).
\bibitem{ZYu07} Z. Zhu and H. Yu, Phys. Lett. B {\bf 645},
 459(2007).
\bibitem{Unruh} W.G. Unruh, Phys. Rev. D {\bf 14}, 870(1976).
\bibitem{Hartle-Hawking}J.~Hartle and S.~Hawking, Phys. Rev. {\bf
D13}, 2188 (1976).

\bibitem{YuZhou07} H. Yu and W. Zhou, Phys. Rev. D {\bf 76}, 027503 (2007), arXiv:0706.2207.

\bibitem{Dicke} R. H. Dicke, Phys. Rev. {\bf93}, 99 (1954).

\bibitem{Dewitt75} B. S. DeWitt, Phys. Rep.  {\bf 19}, 295(1975).

\bibitem{Fulling77} S. M. Christensen and S. A. Fulling, Phys. Rev. D {\bf15}, 2088
(1977).

\bibitem{Candelas80} P. Candelas, Phys. Rev. D {\bf21}, 2185 (1980).

\bibitem{Tolman}R.~Tolman, Phys. Rev. {\bf 35}, 904 (1930); \\
R.~Tolman and P.~Ehrenfest, Phys. Rev. {\bf 36}, 1791 (1930).

\end{thebibliography}
\end{document}